# Providing Local Content in a Hybrid Single Frequency Network using Hierarchical Modulation

Hong Jiang, Paul Wilford and Steve Wilkus

*Abstract* — A hierarchical modulation method is proposed for providing local content in a hybrid satellite and terrestrial single frequency network such DVB-SH. The hierarchical modulation is used to transmit both global and local content in terrestrial transmitters. The global content is transmitted with high priority layer of the hierarchical modulation, and the local content is modulated with the low priority layer of the hierarchical modulation. The satellite transmits global content only. The performance of the hierarchical system for both global and local content is analyzed.

## I. INTRODUCTION

The DVB-SH standard [2] specifies a hybrid satellite and terrestrial single frequency network (SFN) to provide digital service to mobile terminals. In such a hybrid SFN, the service is primarily provided by the signal from an inclined orbit satellite. Low power terrestrial transmitters are deployed as gap fillers in areas where there is blockage of the satellite signal. Both the satellite and terrestrial transmitters broadcast the same waveform in the same frequency band. Synchronization of time and frequency [6] is provided to ensure proper reception in overlapping regions covered by both satellite and terrestrial transmitters. Since the satellite and the terrestrial transmitters broadcast the same waveform, and the satellite covers the entire service area, it is challenging for the hybrid SFN to efficiently provide local content – services that are of interest to particular geographic locations only, such as traffic and weather reports, local news and advertisements. In many existing satellite or hybrid broadcast systems, each local programming is carried throughout the entire network, even to the regions where the programming is of no interest. This arrangement results in significant waste of bandwidth.

Hierarchical modulation can be used as an efficient mechanism to provide local services. Hierarchical modulation has been adopted by the DVB-T/H/SH standards [1], [2]. Using hierarchical modulation to provide local service in a multi-frequency network (MFN) has been considered previously [3]. The purpose of this paper is to propose a mechanism to provide local content in a hybrid satellite and terrestrial SFN. The proposed system is a hierarchical modulation system in which the local content is transmitted by the low priority bit stream. Analysis will be made to evaluate the performance of the system for both local and global content.

The organization of the paper is as follows. In section II, the proposed system is described. Section III makes comparison of hierarchical system with QPSK system, and establishes a theoretical framework to analyze the performance of the system. In Section IV, two configurations of a hybrid system with hierarchical modulation are given, and their performances are analyzed.

## II. THE PROPOSED SYSTEM

The baseline system is a hybrid satellite and terrestrial single frequency network, in which the QPSK constellation is used for both satellite and terrestrial transmissions. The baseline system provides global content which is broadcast to all mobile terminals in the service area covered in the network.

The proposed system uses a hierarchical modulation for the terrestrial transmitters in the baseline system. In addition to broadcasting the global content of the baselines system, the proposed hierarchical system also provides local content from terrestrial transmitters. Local content is added to terrestrial transmitters as needed.

The terrestrial transmitters that have overlapping coverage area form a cluster. Local content in a cluster is targeted to mobile terminals within the coverage area of the cluster. Local content transmitted by transmitters within a cluster must be the same, but the local content may be different between clusters. Different clusters may provide different local content.

The satellite broadcasts global content only. The parameters of the satellite transmission are the same as in the baseline system. A terrestrial transmitter may transmit only the global content, or both the global and local content. A terrestrial transmitter providing only the global content uses the QPSK modulation as in the baseline system. A terrestrial transmitter providing both the global and local content uses a 16-QAM hierarchical modulation.

In a hierarchical modulation, the constellation is allowed to be non-uniform. A hierarchical modulation carries two separate and independent bit streams. The high priority (HP) bit stream is modulated with constellation separated by the quadrants, and the low priority (LP) bit stream is modulated by the constellation within each quadrant. In the DVB standards, [1] [2], the hierarchical parameter $\alpha$ is defined as the minimum distance separating two constellation points carrying different HP bit values divided by the minimum

---

Hong Jiang and Paul Wilford are with Bell Labs, Alcatel-Lucent, 700 Mountain Ave, Murray Hill, NJ 07974, USA. Steve Wilkus is with Bell Labs, Alcatel-Lucent, 791 Holmdel-Keyport Rd, Holmdel, NJ 07733, USA.

distance separating any two constellation points. As shown in Figure 1, if the minimum distance between two constellation points from two different quadrants is $a$, and the minimum distance between two constellation points within a quadrant is $b$, then the hierarchical parameter $\alpha = \dfrac{a}{b}$.

In the proposed system, the HP bit stream is used to transmit global content and the LP bit stream is used to transmit local content. Figure 1 illustrates the hierarchical modulation for the global and local content.

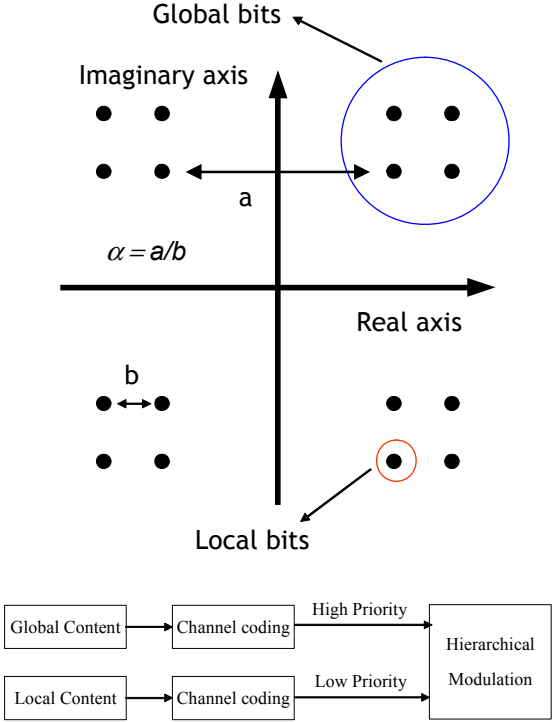

**Figure 1. Hierarchical modulation for global and local content**

The global content and local content may use separate channel coding configurations. The error correction code and code rate of the global content must be the same as those in the baseline system.

Terrestrial transmitters in a cluster must use the same hierarchical parameter $\alpha$ and the same error correction code and code rate. Transmitters in different clusters may choose different hierarchical parameters, and/or different error correction codes and code rates for the local content in LP. For example, transmitters in one cluster may use the convolutional code of code rate 1/3 for the local content, while transmitters in another cluster may use Turbo code of rate 2/3 for their local content.

Figure 2 illustrates the proposed hybrid hierarchical satellite and terrestrial single frequency network capable of providing local content.

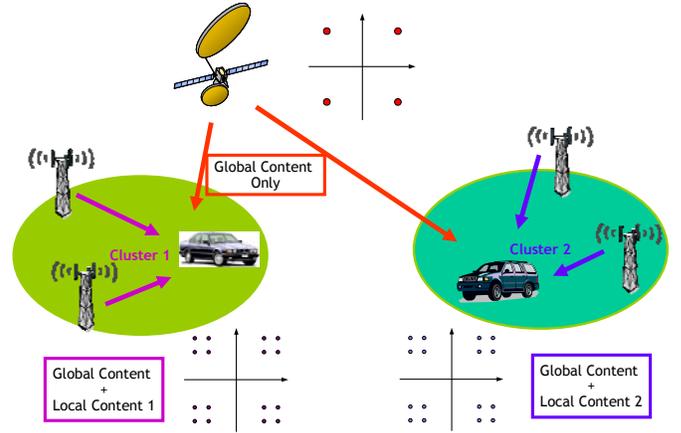

**Figure 2. The proposed hierarchical system**

### 2.1. Hierarchical parameter

The 16-QAM hierarchical modulation for a terrestrial transmitter is configured by the hierarchical parameter $\alpha$. The DVB-SH standard allows the values of the hierarchical parameter $\alpha$ to be 1, 2 or 4 [2]. In this proposal, $\alpha$ may be an arbitrary real number $\geq 1$, including non-integers. The terrestrial transmitters in the same network need not to use a same value for $\alpha$; non-overlapping terrestrial transmitters belonging to different clusters may use different values for $\alpha$. This provides flexibility for terrestrial transmitters to transmit local content with different bit rate and error protection to meet the link budget of each individual cluster.

In this proposal, the value of $\alpha$ is not transmitted explicitly, because there may not be a unique $\alpha$ in the network. Instead, the hierarchical parameter $\alpha$ is embedded in the modulation to be described later. This requires an extension of hierarchy information in the TPS signal of the DVB-SH standard. When the TPS signaling indicates non-hierarchical transmission, this should be interpreted in the new proposed system as either non-hierarchical transmission, or hierarchical transmission with a varying $\alpha$ in which the receiver needs to detect the presence of local content using the modified pilots described below. The detection of local content will be discussed in Section III. Both satellite and terrestrial transmitters in the same network with varying $\alpha$ must transmit the same special value in the hierarchy information of the TPS signal.

The value of $\alpha$ may be embedded in the modulation by using the pilots in the DVB-SH OFDM symbols. For a given subcarrier location where a pilot is located, let the pilots in the existing DVB-SH standard be ordered as $P_i, i = 0,1,2,...$ The index $i$ is the index of OFDM symbols, incrementing with time over one subcarrier. These pilots are BPSK modulated in the existing DVB-SH. For continual pilots, which appear at the same subcarriers all the time, the proposed modified pilots will be modulated as

$$\widetilde{P}_i = P_i + j(-1)^i \frac{P_i}{\alpha+1}, \quad i = 0,1,2,..., \quad j = \sqrt{-1}.$$

For scattered pilots, which are located at different subcarriers in different OFDM symbols, the pilots will be modified as follows. First, if an existing scattered pilot appears at subcarrier $k$ with OFDM symbol $i$, the same subcarrier of OFDM symbol $i+1$ will be used as a modified pilot. In this way, a modified pilot will always appear twice (in time) at the same subcarrier no matter whether it is a continual or scattered pilot. Secondly, the two scattered pilots located at subcarrier $k$ with OFDM symbols $i$ and $i+1$ are modified as

$$\widetilde{P}_i = P_i + j\frac{P_i}{\alpha+1}, \quad \widetilde{P}_{i+1} = P_i - j\frac{P_i}{\alpha+1}.$$

Note that in above definitions, the subcarrier index $k$ is omitted. The design of the modified pilots guarantees not only that a modified pilot of any subcarrier location appears in at least two OFDM symbols, but also that the vertical components of the two consecutive modified pilots have opposite polarity. This design will assist channel estimate to be discussed later.

The modified pilots are illustrated in Figure 3.

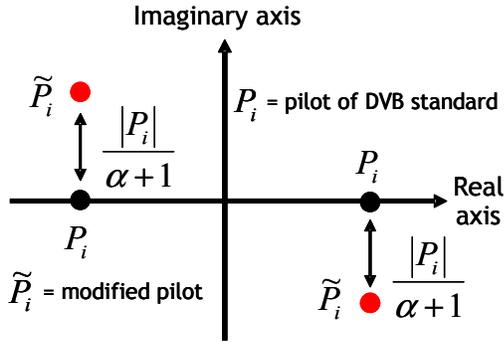

Figure 3. Modified pilots

The distance between a modified pilot and the original pilot is exactly the distance between two constellation points in the same quadrant of the 16-QAM modulation divided by the minimum distance between the four centers, each of which is formed by taking the average of the four constellation points within a quadrant. The modified pilots provide a mechanism for a receiver to detect the hierarchical information. The modulation of TPS signal can be modified similarly as shown in Figure 3. The modified TPS signal is modulated so that information such as the code rate for the LP bit stream is transmitted with the modified TPS signal. Adding of information such as code rate for the LP bit stream into the TPS signal is similar to adding local content to the global QPSK constellation. The added information will impose a penalty to TPS reception, similar to the penalty on the global content due to the insertion of local content.

The modified pilots also impose a penalty on transmission power because they carry higher energy than the original pilots. The power of the modified pilots is increased by a factor of $1 + \frac{1}{(\alpha+1)^2}$. This additional power is in line with the increased power of other subcarriers incurred from the insertion of local content. An analysis of the impact of the increased power will be given in the next section. A further penalty on bandwidth efficiency is incurred because an additional subcarrier is used for each scattered pilot.

The modified pilots are transmitted only in terrestrial transmitters carrying local contents.

The satellite signal is always transmitted with $\alpha = +\infty$, which corresponds to the QPSK modulation. A terrestrial transmitter may transmit QPSK signal (with $\alpha = +\infty$), if no local content is needed. When local content is added, the transmitter uses 16-QAM hierarchical modulation, with a finite value of $\alpha \geq 1$.

### 2.2. Receivers

In order to receive the local content, a receiver must be capable of demodulating a signal with hierarchical modulation as described in section 2.1.

The modified pilots, Figure 3, are used for the receivers to determine whether the received signal is hierarchically modulated, and if yes, what the value of $\alpha$ is. When a hierarchical modulation is detected, the global and local content are extracted from the HP bit stream and the LP bit stream, respectively, and they are separately decoded with possible different error correction codes and code rates.

When no hierarchical modulation is detected, either due to the signal being QPSK (transmitted by the satellite, for example), or due to the inability to detect the 16-QAM (because $\alpha$ is too large, or the signal is too noisy due to a weak signal power), only the QPSK signal will be demodulated, resulting in global content only.

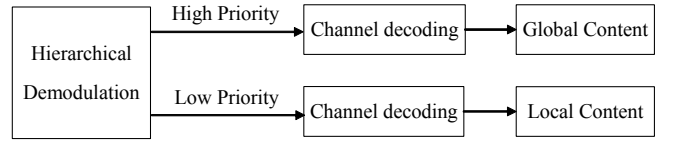

Figure 4. Hierarchical receiver for global and local content

A receiver needs to detect whether local content is present and demodulate the local content automatically without user configuration. The detection is assisted with the modified pilots. A receiver can be designed to extract the global and local content as described in [3]. The receiver of [3] extracts the global content from the received signal first, and then extracts the local contents by using the decoded global content. [3] also proposed an iterative scheme in which after the local content is extracted, the local content is further used to improve the decoding of the global content, which is in turn used to improve the decoding of local content and so on.

## III. PERFORMANCE ANALYSIS

Performance comparison of the proposed hierarchical system with the baseline system will be made in this section.

The parameter $\alpha$ controls the error characteristics of bits in the global content and local content. A large value of $\alpha$ provides more reliable transmission channel for global content, but a more noisy channel for the local content. A small value of $\alpha$, on the other hand, reduces reliability of the global content, but increases the robustness of the local content. A detailed analysis of the tradeoff between the global content and local content with different values of $\alpha$ can also be found in [3].

In a region covered by the satellite only, there is only a QPSK signal. The proposed hierarchical system behaves exactly the same as the baseline system (QPSK modulation), and only the global content is available.

Near a terrestrial transmitter with local content transmission, where the terrestrial signal dominates, the system performs as the 16-QAM system with hierarchical modulation. With the same transmitting power as the baseline system, the performance of the global content bits in the hierarchical system will be degraded as compared to the baseline system. The degradation is due to sharing the transmitting power between the global content and local content. The degree of degradation is a function of parameter $\alpha$; the larger $\alpha$ is, the smaller the degradation is, see [3], and the analysis later in this section.

The degradation of the global content as compared to the baseline system can be handled with the following three options:

**a) Do nothing**

In this option, the coverage area of the transmitter in which the global content can be reliably received is reduced. The local content is added at the expense of the reduced coverage of the global content. The amount of degradation will be analyzed in details later in this section.

**b) Increase the error protection of global content**

In this option, the code rate of the global content can be reduced to provide more error protection to the global content, so that the area in which the global content can be reliably received is the same as the baseline system. Consequently, the user data rate of the global content is reduced, and the local content is added at the expense of the reduced bit rate for the global content. A disadvantage of this option is that it affects configuration of entire network, even when the local content is added to only one terrestrial transmitter.

**c) Increase transmitter power**

In this option, the transmitting power is increased to offset the degradation of the global content, so that the global content has the same bit rate and performance as the baseline system. The local content is added at the expense of additional transmitting power. An advantage of this option is that it provides the flexibility to each individual terrestrial transmitter. This makes the expansion, upgrading of the network easy. When local content is needed in a geographic region, only the terrestrial transmitters within that region need to be re-configured or upgraded. The other parts of the network will not be affected in any way. However, it may not always be possible to increase the transmitting power because the baseline system may already be transmitting at the maximum power allowed by government regulations.

In the hybrid region, where both the satellite and terrestrial signals are present with similar strength, the satellite signal, which carries global content only, adds constructively with the global content bits of the terrestrial transmitter, assuming synchronization is properly maintained as in the baseline system. The global content bits in the proposed system will have same benefit of the SFN gain as the baseline system. Therefore, the presence of the satellite signal helps the reception of the global content in the terrestrial transmission.

The presence of the satellite signal has no or little effect on the performance of the local content bits in the terrestrial signal. This is because the adding of the satellite signal to the terrestrial signal does not affect the decision distance of the LP constellation points modulated with the local content bits. This will be further discussed in subsection 3.3.

Next, the performance of each of the global content and local content bits will be analyzed. The analysis will provide guidelines on how to choose network configuration parameters to make trade-off of bitrate/performance between the global content and the local content.

One important measurement of the system performance is the bit error ratio (BER). We are interested in the BER in the global and local content bit streams, with different values of $\alpha$. While it is possible to obtain BER results by simulations, the computation would be very intense if the BER measurements must be made with different values of $\alpha$, different code rates for both global and local content.

On the other hand, the BER results for QPSK in AWGN channel with various code rates of Turbo coding are well known; see, for example, [4] and [5]. In this section, we will develop an analytic tool that will allow us to obtain the BER for both the global and local content with different values of hierarchical distance $\alpha$, based on the simulation results of BER for QPSK modulation. This is achieved by using a concept of effective Es/No.

All analysis will be performed under the assumption of the AGWN channel.

### 3.1. Effective Es/No

In the proposed hierarchical system, the received power is shared by the global content and local content. The total received power, therefore, can be split between them, yielding effective power per symbol (Es) over noise (No) for global content and local content. The performance of the each can be analyzed using the effective Es/No. Figure 5 visualizes the definition of effective Es/No.

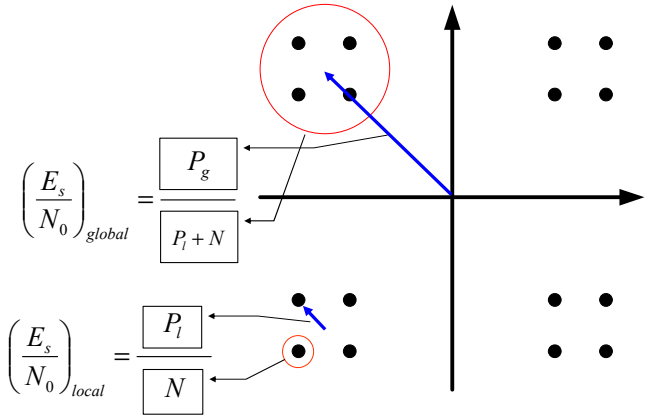

**Figure 5 Definition of effective Es/No**

The total received power can be expressed as the sum of the powers for global content and local content as given blow

$$P_r = P_g + P_l, \qquad (1)$$

where $P_r$ is the total received power, $P_g$ is the power for global content, and $P_l$ is the power for local content.

Let $N$ be the power of Gaussian noise, the carrier to noise ratio (C/N) at the receiver is therefore given by

$$CNR = \frac{P_r}{N} = \frac{P_g + P_l}{N} = \frac{P_g}{N} + \frac{P_l}{N} = \frac{P_g}{P_l + N}\frac{P_l + N}{N} + \frac{P_l}{N}.$$

The power of local content constitutes an additional noise to the global content, and therefore, the effective energy to noise ratio for global content is

$$G = \left(\frac{E_s}{N_0}\right)_{global} = \frac{P_g}{P_l + N}. \qquad (2)$$

We note that the noise represented by the local content is not Gaussian distributed. Strictly speaking, the noise in the denominator of (2) is not Gaussian distributed, but it can nevertheless be sufficiently approximated in practice by a Gaussian noise. The simulation results presented later also support this assumption.

The effective energy to noise ratio for local content is simply

$$L = \left(\frac{E_s}{N_0}\right)_{local} = \frac{P_l}{N}. \qquad (3)$$

With these definitions, the received C/N is given by

$$\begin{aligned} CNR &= \frac{P_g}{P_l + N}\frac{P_l + N}{N} + \frac{P_l}{N} \\ &= G + L + G \cdot L \\ &= \left(\frac{E_s}{N_0}\right)_g + \left(\frac{E_s}{N_0}\right)_l + \left(\frac{E_s}{N_0}\right)_g \cdot \left(\frac{E_s}{N_0}\right)_l. \end{aligned} \qquad (4)$$

In terms of the hierarchical parameter $\alpha$ as defined in Figure 1, we have

$$P_g = 2(1+\alpha)^2, \text{ and } P_l = 2.$$

The ratio of the powers is

$$(1+\alpha)^2 = \frac{P_g}{P_l} = \frac{P_g}{P_l + N}\frac{P_l + N}{P_l} = G\frac{L+1}{L}, \qquad (5)$$

where $G$ and $L$ are given by equations (2) and (3), respectively.

Combining equations (4) and (5), the effective energy to noise ratios are given by

$$G = \left(\frac{E_s}{N_0}\right)_{global} = \frac{(1+\alpha)^2}{1 + CNR + (1+\alpha)^2} CNR \qquad (6)$$

$$L = \left(\frac{E_s}{N_0}\right)_{local} = \frac{1}{1+(1+\alpha)^2} CNR. \qquad (7)$$

In above equations, $CNR$ is the carrier to noise ratio (C/N) of the hierarchically modulated signal at the receiver.

Equations (6) and (7) give the effective Es/No for global content and local content as functions of the received C/N. These functions are plotted for various values of $\alpha$ as shown in Figure 6.

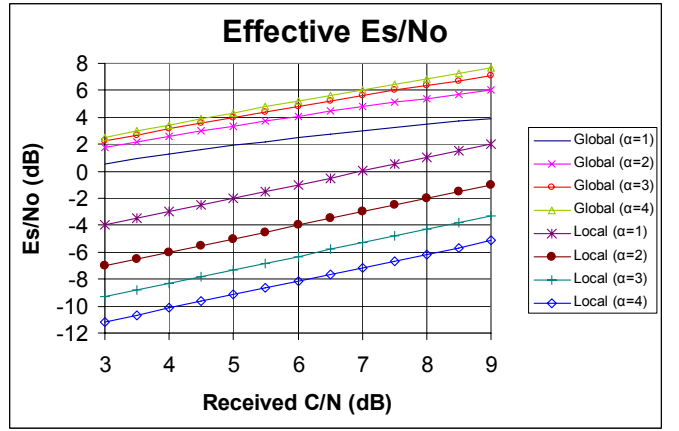

**Figure 6. Effective Es/No**

The effective Es/No can be used to derive BER of global and local content as a function of C/N in the received signal.

### 3.2. C/N requirements for global and local content

In the proposed hierarchical system, each of the global content and local content is effectively modulated with QPSK. For a given C/N of the received hierarchical signal, the effective

Es/No of the QPSK for each of the global and local content can be calculated by (6) and (7), respectively, as shown in Figure 6. The BER curve of QPSK as a function of Es/No in AWGN is well known, and with the help of it, BER of the global and local content can be computed for various combinations of code rates, values of $\alpha$ and C/N.

For example, in order to find the BER of the global bits in AWGN for given C/N and $\alpha$, equation (6) is used to compute the effective Es/No for the given C/N and $\alpha$. Then the resulted Es/No is used to find the corresponding BER in the BER vs Es/No curve of the QPSK modulation. Similarly, the BER for local content is found by using equation (7).

Table 1 shows the required C/N in dB to achieve error rate of BER ≤ $10^{-5}$ in AWGN channel for various code rates using the Turbo code as defined in DVB-SH. The required C/N values in column 2 for QPSK are quoted from the DVB-SH implementation guideline [5]. Columns 3 and 4 are the required C/N values for the hierarchical 16-QAM ($\alpha$ = 2) to achieve the same BER performance. The required C/N values in column 3 for global content are derived from equation (6) and column 2. To find the required C/N for a given code rate, use the corresponding value in column 2 (QPSK) as *G* in equation (6), and solve for *CNR* from it (with α = 2). Similarly, the required C/N values in column 4 for local content are derived from equation (7) and column 2.

**Table 1 Required C/N (dB) to achieve BER ≤ $10^{-5}$**

| Code Rate | C/N (dB) for QPSK (Ref. [5]) | C/N (dB) for Hierarchical 16QAM This paper (α=2) | |
|---|---|---|---|
| | | Global | Local |
| 1/5 | -3.6 | -3 | 6.4 |
| 2/9 | -3.1 | -2.4 | 6.9 |
| 1/4 | -2.5 | -1.8 | 7.5 |
| 2/7 | -1.8 | -1 | 8.2 |
| 1/3 | -0.9 | 0 | 9.1 |
| 2/5 | 0.1 | 1.1 | 10.1 |
| 1/2 | 1.4 | 2.6 | 11.4 |
| 2/3 | 3.5 | 5.2 | 13.5 |

The theoretical results for the hierarchical 16-QAM listed in Table 1 match very well with published simulation results. A comparison of the results obtained in this paper with simulation results in [7] is given in Table 2, which shows the required C/N values in dB for the hierarchical 16-QAM to achieve error rate of BER ≤ $10^{-5}$ in AWGN channel. Among all code rates, in both the global and local content, the maximum difference between the theoretical results of this paper and the simulation results of [7] is 0.2dB.

**Table 2 Comparison of required C/N for BER ≤ $10^{-5}$**

| Code Rate | C/N (dB) for Hierarchical 16QAM This paper (α=2) | | C/N (dB) for Hierarchical 16QAM Simulation [7] (α=2) | |
|---|---|---|---|---|
| | Global | Local | Global | Local |
| 1/5 | -3 | 6.4 | -2.9 | 6.5 |
| 2/9 | -2.4 | 6.9 | -2.3 | 7.1 |
| 1/4 | -1.8 | 7.5 | -1.7 | 7.6 |
| 2/7 | -1 | 8.2 | -1 | 8.3 |
| 1/3 | 0 | 9.1 | -0.1 | 9.1 |
| 2/5 | 1.1 | 10.1 | 1.1 | 10.2 |
| 1/2 | 2.6 | 11.4 | 2.7 | 11.6 |
| 2/3 | 5.2 | 13.5 | 5.2 | 13.7 |

It is also possible to plot the curves of BER vs C/N for various values of hierarchical distance $\alpha$, by using the BER curve for the QPSK modulation and the process described earlier. Examples will be given in the next section.

### 3.3. Effect of satellite signal on the terrestrial global and local content

In this subsection, we analyze how the satellite signal, consisting of global content only, will affect the global and local content of the terrestrial signal in a hybrid region where both the satellite and terrestrial signals are present.

Denote by $S_k$ a QPSK constellation at subcarrier $k$ in the transmitter. Then the transmitted frequency-domain sequence for the global content can be written as $S_k^g$. As shown in Figure 5, the transmitted frequency-domain sequence for the global and local content can be written as $S_k^g + \frac{1}{1+\alpha} S_k^l$. In above, $S_k^g$ and $S_k^l$ denote the QPSK constellations at subcarrier $k$, carrying global and local content, respectively. $S_k^g$ and $S_k^l$ have the same amplitude. The size of the local content constellation is reflected through the factor $\frac{1}{1+\alpha}$ in front of $S_k^l$.

Let $R_k^s$ be the received frequency-domain sequence of the satellite signal, which contains global content only. Let $R_k^t$ be the received frequency-domain sequence of the terrestrial signal, which contains both global and local content. Then we have

$$R_k^s = A_k^s S_k^g + N_k^s, \tag{8}$$

$$R_k^t = A_k^t \left( S_k^g + \frac{1}{1+\alpha} S_k^l \right) + N_k^t. \tag{9}$$

In (8) and (9), $A_k^s$ and $A_k^t$ are complex values representing the channel responses of the satellite and terrestrial paths,

respectively. $N_k^s$ and $N_k^t$ are additive noise. When the local content is not transmitted, $\alpha = +\infty$. In the hybrid region where both satellite and terrestrial signals are present, the received signal can be written as

$$R_k^h = (A_k^s + A_k^t)S_k^g + \frac{A_k^t}{1+\alpha}S_k^l + N_k^h,$$

or equivalently,

$$R_k^h = A_k^g S_k^g + A_k^l S_k^l + N_k^h, \qquad (10)$$

where $N_k^h$ is the additive noise, and

$$A_k^g = A_k^s + A_k^t, \ A_k^l = \frac{A_k^t}{1+\alpha}. \qquad (11)$$

Equation (10) shows that in the received hybrid signal, the relative geometry of the constellations for the global and local content is, in general, no longer the same as that depicted in Figure 1 because $A_k^g$ and $A_k^t$ may have different phases and amplitudes. Nevertheless, demodulation of the global and local content can be achieved after the channel responses $A_k^g, A_k^l$ are estimated by using the modified pilots. Channel estimation will be described in the next subsection.

First, we consider the effect of the satellite signal on the performance of the global content in the terrestrial signal. We rewrite the received terrestrial signal of (9) as

$$R_k^t = A_k^t S_k^g + A_k^l S_k^l + N_k^t. \qquad (12)$$

The first term on the right hand side (RHS) of (12) is the global content constellation, and the second and third terms on the RHS of (12) represent noise to the global content, whose effect has been studied in the previous subsections. Comparing (12) to (10), we find that the noise to the global content is the same in the terrestrial signal (12) and the hybrid signal (10), satellite signal plus terrestrial signal, but the global content constellations are different, $A_k^t S_k^g$ for the terrestrial signal, and $A_k^g S_k^g$ for the hybrid signal. When the satellite propagation path and the terrestrial propagation path are independent, which is a reasonable assumption in reality, we have

$$\begin{aligned}\mathbf{E}(|A_k^g|^2) &= \mathbf{E}(|A_k^t + A_k^s|^2) \\ &= \mathbf{E}(|A_k^t|^2) + \mathbf{E}(|A_k^s|^2) \geq \mathbf{E}(|A_k^t|^2).\end{aligned} \qquad (13)$$

In (13), $\mathbf{E}(\cdot)$ is the expected value. Equation (13) shows that, on average, the constellation $A_k^g S_k^g$ of the hybrid signal has larger decision distance than $A_k^t S_k^g$ of the terrestrial signal. Therefore, the global content in the hybrid signal has a better performance than that in the terrestrial signal alone. The performance is gained by a factor of

$$\frac{\mathbf{E}(|A_k^g|^2)}{\mathbf{E}(|A_k^t|^2)} = 1 + \frac{\mathbf{E}(|A_k^s|^2)}{\mathbf{E}(|A_k^t|^2)}. \qquad (14)$$

This performance gain represents the SFN gain due to the availability of both terrestrial and satellite signals. Note that this SFN gain is realized even though the satellite signal, which contains global content only, is different from the terrestrial signal, which contains both the global and local content. As shown in (14), the stronger the satellite signal is (relative to the terrestrial signal), the higher the gain is.

Next, let us discuss the effect of the satellite signal on the local content of the terrestrial signal. For this purpose, we assume the receiver is designed as described [3] in which the global content is first extracted from the received signal, after which the local content is then extracted.

Therefore, to consider the performance of the local content, we assume the global content has been extracted (successfully). When the received signal consists of terrestrial signal only, the extraction of local content is based on the following equation reformulated from (9) or (12)

$$R_k^t - A_k^t S_k^g = A_k^l S_k^l + N_k^t. \qquad (15)$$

In equation (15), the left hand side (LHS) is assumed to be known because the global content has been decoded. We will see how to estimate channel responses $A_k^t$ and $A_k^l$ in the next subsection. The performance of the local content in the terrestrial-only signal is therefore determined by the decision distance of $A_k^l S_k^l$ and the power of noise $N_k^t$. Similarly, in the hybrid signal, the local content is demodulated and decoded based on the following equation reformulated from (10)

$$R_k^h - A_k^g S_k^g = A_k^l S_k^l + N_k^h, \qquad (16)$$

and the performance of the local content in the hybrid signal is determined by the decision distance of $A_k^l S_k^l$ and the power of noise $N_k^h$. A comparison of (15) and (16) shows that the decision distance in both cases is the same, and the noise power is the same too. This concludes that the performance of the local content is the same for both the terrestrial-only and the hybrid signals. The presence of the satellite signal does not degrade the performance of local content in the terrestrial signal.

It is worthwhile to point out that with the implementation of the receiver as described in [3], the presence of the satellite signal in fact helps slightly the performance of the local content. This is because the LHS of (16) may be more accurate than the LHS of (15) due to the SFN gain of the global content as discussed earlier. In other words, since the demodulation and decoding of the local content depend on the successful decoding of the global content, the SFN gain of global content due to the presence of satellite signal makes the successful decoding of local content more likely.

### 3.4. Channel estimation

The channel responses $A_k^g, A_k^l$ can be estimated by using the modified pilots. Although the modified pilots are not QPSK constellations, they can nevertheless be written as

$$S_k^g + \frac{1}{1+\alpha} S_k^l. \tag{17}$$

Referring to Figure 3, $S_k^g$ represents the horizontal vector of the modified pilots, while $S_k^l$ represents the vertical vector having the same amplitude as $S_k^g$. The reason for the factor $\frac{1}{1+\alpha}$ to be in the definition of the modified pilots is so that (17) can represent both the modified pilots as well has the hierarchical modulated constellation consisting of both global and local content. Because of this, the parameter $\alpha$ does not explicitly participate in the receiver processing, as we will see shortly.

At a subcarrier $k$ where a modified pilot is located, the two pilots received in two consecutive OFDM symbols can be written as

$$\begin{aligned}(R_k^h)_1 &= A_k^g (S_k^g)_1 + A_k^l (S_k^l)_1 + (N_k^h)_1 \\ (R_k^h)_2 &= A_k^g (S_k^g)_2 + A_k^l (S_k^l)_2 + (N_k^h)_2\end{aligned}, \tag{18}$$

where the subscripts 1 and 2 signify the two time instances. In (18), we have assumed the channels $A_k^g, A_k^l$ remain relatively constant during the two OFDM symbol periods. This assumption is valid in AWGN channels, or in channels with slow fading. For channels with fast fading, see comments in Section 3.5. From (18), the channel responses can be estimated by solving the following 2x2 linear system of equations for $\widetilde{A}_k^g, \widetilde{A}_k^l$

$$\begin{aligned}\widetilde{A}_k^g (S_k^g)_1 + \widetilde{A}_k^l (S_k^l)_1 &= (R_k^h)_1 \\ \widetilde{A}_k^g (S_k^g)_2 + \widetilde{A}_k^l (S_k^l)_2 &= (R_k^h)_2\end{aligned}. \tag{19}$$

In (19), the quantities with subscript 1 or 2 are known. On the LHS, they are the vector components of the known transmitted modified pilots, and on the RHS, they are the received signals. Equation (19) is non-singular because the vertical vector components of the two consecutive modified pilots ( $(S_k^l)_1$ and $(S_k^l)_2$ ) have opposite polarity. In fact, by the definition of the modified pilots, (19) is equivalent to

$$\begin{aligned}\widetilde{A}_k^g + \widetilde{A}_k^l &= (R_k^h / S_k^g)_1 \\ \widetilde{A}_k^g - \widetilde{A}_k^l &= (R_k^h / S_k^g)_2\end{aligned}, \tag{20}$$

where $(S_k^g)_1$ and $(S_k^g)_2$ are the horizontal components of the two consecutive (in time) modified pilots. Equation (20) can be solved easily. In practice, equation (20) can be formed with the received signals averaged over a certain time period in order to combat noise. That is, instead of using $(R_k^h/S_k^g)_1, (R_k^h/S_k^g)_2$, the RHS of (20) is replaced by $avg((R_k^h/S_k^g)_1), avg((R_k^h/S_k^g)_2)$. The time interval of average should be chosen so that the channels $A_k^g, A_k^l$ remain relatively constant during the averaging period.

Note that the parameter $\alpha$ does not need to be estimated explicitly. It is estimated as part of the channel response $\widetilde{A}_k^l$.

It can be easily verified that, in the absence of noise, the solution $\widetilde{A}_k^l = 0$ is obtained if no local content is transmitted ( $\alpha = +\infty$ ), or if the received signal contains satellite signal only. Therefore, in the receiver implementation, local content is detected by examining the amplitude of the estimated $\widetilde{A}_k^l$. A small amplitude of $\widetilde{A}_k^l$ is interpreted as either 1) no local content is transmitted, or 2) the local content is too weak to be detected, which is the case when, for example, the receiver is too far away from the terrestrial transmitter of the local content.

Finally, we point out that the accuracy of the channel estimate only depends on the power of noise $N_k^h$. In particular, the accuracy in the estimates $\widetilde{A}_k^g, \widetilde{A}_k^l$ is independent of the power level of the satellite component within the hybrid signal. This is readily verified by looking at the errors in the estimates

$$\Delta_k^g = \widetilde{A}_k^g - A_k^g, \Delta_k^l = \widetilde{A}_k^l - A_k^l. \tag{21}$$

Subtracting (18) from (19), and applying the definition of the modified pilots, we find the errors in channel estimates satisfy

$$\begin{aligned}\Delta_k^g + \Delta_k^l &= (N_k^h / S_k^g)_1 \\ \Delta_k^g - \Delta_k^l &= (N_k^h / S_k^g)_2\end{aligned}. \tag{22}$$

Equation (22) shows that the errors $\Delta_k^g, \Delta_k^l$ are functions of the noise $N_k^h$ only.

### 3.5. Effect of non-AWGN channels

The analysis performed above is under the assumption of AWGN channel, and it has not taken into consideration of multipath propagation and mobile reception. There has been study [8] showing that MER degradation exists because of the combination of different propagation paths in an SFN, and that an increase of the C/N due to the co-existence of different rays may not be beneficial for the reception process [8]. In addition, if signal fading is fast enough so that there is a significant channel change between two consecutive OFDM symbols, the channel estimate of section 3.4 may not be adequate. As a result, the more challenging reception conditions of a mobile network may degrade the performance of the proposed system as predicted by the analysis. Physical layer simulations and field experiments are necessary tools to aid the design of such a system for a mobile network.

## IV. EXAMPLES OF THE PROPOSED SYSTEM

Two examples of system configuration to provide local content in a hybrid hierarchical modulation SFN will be given in this section. In the first example, the value of $\alpha$ is limited to those allowed by the existing DVB-SH standard, and in the second example, a non-integer value of $\alpha$ is used.

In both examples, the baseline QPSK system is a DVB-SH system with the following parameters:

- Bandwidth = 5MHz
- FFT size = 2K
- GI length = 1/8
- Turbo coding code rate = 2/3
- User data rate (MPEG) = 4.937 Mbps
- Targeted BER after Turbo coding = $10^{-5}$ in AWGN

The baseline system with these parameters will also be referred to as Reference Configuration (RC) in the rest of this section.

Local content will be added to the baseline system using a hierarchical modulation with parameter $\alpha$. The desired user data rate of the local content is 1.55 Mbps. The code rate of 2/9 for the local content supports a user data rate of more than 1.64 Mbps. Therefore, in both examples, the code rate for the local content in LP will be 2/9. The global content in HP has the same code rate as the baseline system.

In both examples, the terrestrial transmitters with local content maintain the same transmission power as the baseline system. Therefore, the addition of the local content will degrade the global content as compared to the baseline system. Analysis will be made to assess the degree of degradation, and the configurations will be made to minimize the degradation, to the extent of meeting other constraints. All analysis will be performed under the assumption of the AGWN channel.

### 4.1. Example Configuration 1

In the first configuration, the hierarchical parameter $\alpha$ will be limited to the integer values allowed by the DVB standards [1], [2], namely, $\alpha = 1, 2, 4$. The goal is to choose a standardized $\alpha$ to make the trade-off in minimizing the degradation of the global content and maximizing the coverage area of local content, the area in which the local content bits have the performance criterion of BER $\leq 10^{-5}$. The parameter $\alpha = 2$ is the optimal among the values allowed by the standards [1], [2]. The details of the system parameters are given in Table 3.

**Table 3 Parameters for configuration 1**

| Design Parameters | RC | Hierarchical Modulation | |
| --- | --- | --- | --- |
|  |  | Global | Local |
| Hierarchical distance (alpha) | N/A | 2 | 2 |
| Code rate | 2/3 | 2/3 | 2/9 |
| User data rate (Mbps) | 4.937 | 4.937 | 1.646 |
| Req'ed C/N (dB) @ BER $10^{-5}$ | 3.5 | 5.2 | 6.9 |

In this configuration, for the global content bits to achieve the BER requirement of BER $\leq 10^{-5}$ in AWGN, the required C/N is 5.2dB as compared to 3.5dB needed in the baseline system to achieve the same performance. Therefore, the global content in the hierarchical system represents an 1.7dB degradation. For the local content bits to achieve the BER requirement of BER $\leq 10^{-5}$ in AWGN, the required C/N is 6.9dB, which is 3.4dB higher than the baseline system.

The BER curves of the baseline system, global content bits and local contents can be derived by using (6), (7) and the simulation results for QPSK from [4]. The BER curves for the current configuration of the hierarchical system are shown in Figure 7.

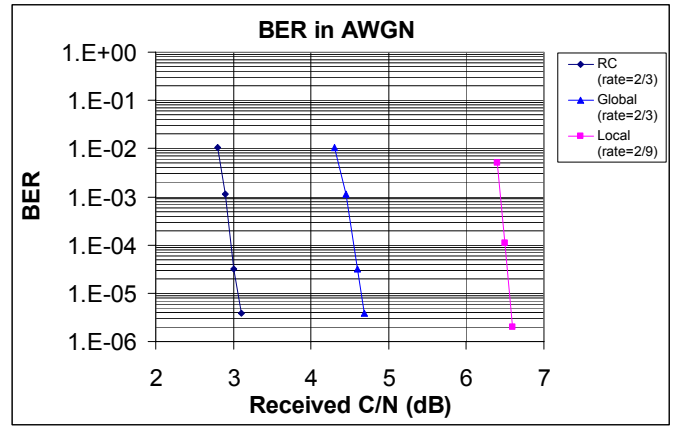

**Figure 7. BER of baseline system, global content and local content for configuration 1**

### 4.2. Example Configuration 2

In this configuration, the hierarchical parameter $\alpha$ will not be limited to the integer values allowed by the DVB standards [1], [2]. The goal is to choose a value of $\alpha$ so that both global and local content bits have the same BER performance. In other words, the goal is for both the global content and local content to have the same coverage area. The parameter $\alpha = 1.6$ meets this goal. The details of the system parameters are given in Table 4.

**Table 4 Parameters for configuration 2**

| Design Parameters | RC | Hierarchical Modulation | |
| --- | --- | --- | --- |
|  |  | Global | Local |
| Hierarchical distance (alpha) | N/A | 1.6 | 1.6 |
| Code rate | 2/3 | 2/3 | 2/9 |
| User data rate (Mbps) | 4.937 | 4.937 | 1.646 |
| Req'ed C/N (dB) @ BER $10^{-5}$ | 3.5 | 5.85 | 5.8 |

As before, the required C/N for the baseline system to achieve BER = $10^{-5}$ in AWGN is 3.5dB. As compared with example configuration 1, the value of $\alpha$ is decreased in this configuration. Consequently, the performance of global bits is degraded as that of local bits is improved. The required C/N for global and local bits to achieve BER = $10^{-5}$ in AWGN are about 5.85dB and 5.8dB, respectively, as shown in Table 4.

The BER curves for the baseline system, the global content and local content of the hierarchical system as functions of C/N of the received signal are shown in Figure 8.

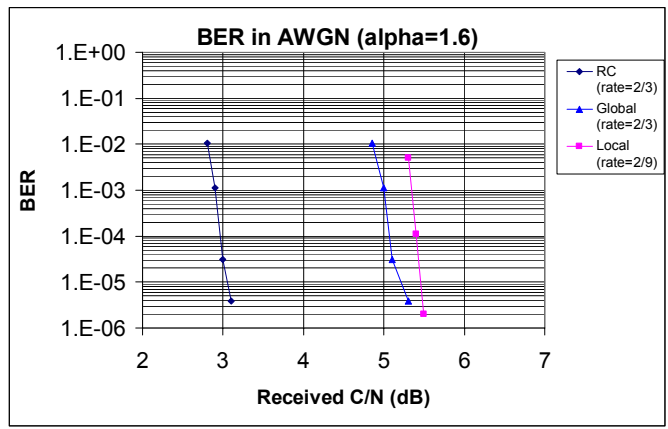

**Figure 8 BER of baseline system, global content and local content for configuration 2**

The BER curves are derived from simulation results for QPSK in [4] and equations (6) and (7).

### V. CONCLUSION

Hierarchical modulation can be used to provide local content efficiently in a satellite/terrestrial hybrid single frequency network. Together with the error correction code rate, the hierarchical parameter $\alpha$ can be adjusted to meet performance requirements of a system, such as bit rate of local content. The proposed hierarchical modulation system is also flexible in network planning and network expansion by allowing different system configuration parameters to be chosen for different terrestrial transmitters in the same network, according to link budget requirement of different geographic regions.

### ACKNOWLEDGEMENT
The authors would like to thank Dmitry Chizhik and Jonathan Ling of Alcatel-Lucent for sharing their simulation results. The authors also thank Gang Huang of Alcatel-Lucent for insightful discussions leading to the improvement of subsections 3.3 and 3.4, and the anonymous reviewers for their valuable comments.